%% file: C-MaStar.tex
\documentclass[fleqn,usenatbib]{mnras}

\usepackage{newtxtext,newtxmath}
\usepackage[T1]{fontenc}
\usepackage{ae,aecompl}
\usepackage{newtxtext,newtxmath}
\usepackage{graphicx}	% Including figure files
\usepackage{amsmath}	% Advanced maths commands
\usepackage{nccmath}
\usepackage{multicol}        % Multi-column entries in tables
\usepackage{bm}		% Bold maths symbols, including upright Greek
\usepackage{pdflscape}	% Landscape pages
\usepackage{orcidlink}
\usepackage{rotating}
\usepackage{tikz}
\usepackage{xurl}
\usepackage{xcolor}
\usepackage{rotating}   % for landscape table
\usepackage{longtable}
\newcommand{\gapprox}{\,\rlap{\lower 2.5pt % > ungefaehr =
\hbox{$\sim$}}\raise 1.5pt\hbox{$>$}\,}
\newcommand{\gsim}{\,\rlap{\lower 2.5pt % > ungefaehr =
\hbox{$\sim$}}\raise 1.5pt\hbox{$>$}\,}
\newcommand{\lapprox}{\,\rlap{\lower 2.5pt % < ungefaehr =
\hbox{$\sim$}}\raise 1.5pt\hbox{$<$}\,}
\newcommand{\lsim}{\,\rlap{\lower 2.5pt % < ungefaehr =
\hbox{$\sim$}}\raise 1.5pt\hbox{$<$}\,}

\def\eeq{\end{equation}}
\def\beq{\begin{equation}}

\title[C-MaStar]{Carbon- and Oxygen-rich stars in MaStar: identification and classification}
\author[Hill et al.]{Lewis Hill$^{1}$\thanks{E-mail:lewis.hill@port.ac.uk}\orcidlink{0000-0001-8052-8080},
Claudia Maraston$^{1}$\thanks{E-mail:claudia.maraston@port.ac.uk}\orcidlink{0000-0001-7711-3677},
Daniel Thomas$^{1,2}$\orcidlink{0000-0002-6325-5671},
Renbin Yan$^{3}$\orcidlink{0000-0003-1025-1711},
Yanping Chen$^{4}$,
\newauthor
Guy S. Stringfellow$^{5}$\orcidlink{0000-0003-1479-3059},
Richard R. Lane$^{6}$\orcidlink{0000-0003-1805-0316},
Jos\'e G. Fern\'andez-Trincado$^{7}$
\\
$^{1}$ Institute of Cosmology, University of Portsmouth, Burnaby Road, Portsmouth PO1 3FX, UK \\
$^{2}$ School of Mathematics and Physics, University of Portsmouth, Lion Gate Building, Portsmouth, PO1 3HF, UK\\
$^{3}$ Department of Physics, The Chinese University of Hong Kong, Shatin, N.T., Hong Kong, China\\
$^{4}$ New York University Abu Dhabi, Abu Dhabi, P.O. Box 129188, United Arab Emirates\\
$^{5}$ Center for Astrophysics and Space Astronomy, University of Colorado, 389 UCB, Boulder, CO 80309-0389, USA\\
$^{6}$ Centro de Investigación en Astronomía, Universidad Bernardo O'Higgins, Avenida Viel 1497, Santiago, Chile\\
$^{7}$ Instituto de Astronom\'ia, Universidad Cat\'olica del Norte, Av. Angamos 0610, Antofagasta, Chile}

\date{Accepted 2024 March 27. Received 2024 March 27; in original form 2022 September 22}

\pubyear{2024}

\begin{document}
\label{firstpage}
\maketitle

\begin{abstract}
Carbon- and Oxygen-rich stars populating the Thermally-Pulsing Asymptotic Giant Branch (TP-AGB) phase of stellar evolution
are relevant contributors to the spectra of $\sim 1$ Gyr old populations. Atmosphere models for these types are uncertain, due to
complex molecules and mass-loss effects. Empirical spectra are then crucial, but samples are small due to the short ($\sim 3$ Myr)
TP-AGB lifetime. Here we exploit the vastness of the MaNGA Stellar library MaStar ($\sim 60,000$~spectra) to identify
C,O-rich type stars. We define an optical colour selection with cuts of $(g-r)>2$~and $(g-i)<1.55(g-r)-0.07$,
calibrated with known C- and O- rich spectra. This identifies C-,O-rich stars along clean, separated sequences. An analogue selection is found in $V,R,I$~ bands. Our equation identifies C-rich and O-rich spectra with predictive performance metric F1-scores of
0.72 and 0.74 (over 1), respectively. We finally identify 41 C- and 87 O-rich type AGB stars in MaStar,
5 and 49 of which do not have a SIMBAD counterpart. We also detect a sample of non-AGB, dwarf C-stars. We further design a
fitting procedure to classify the spectra into broad spectral types, by using as fitting templates empirical C and O-rich spectra. We find remarkably good fits for the majority of candidates and categorise them into C- and O-rich bins following existing
classifications, which correlate to effective temperature. Our selection models can be applied to large photometric surveys (e.g.
Euclid, Rubin). The classified spectra will facilitate future evolutionary population synthesis models.
\end{abstract}

\begin{keywords}
stars: AGB and post-AGB galaxies - stars: carbon - stars: fundamental parameters stars: statistics - surveys - galaxies: stellar content
\end{keywords}

%%%%%%%%%%%%%%%%%%%%%%%%%%%%%%%%%%%%%%%%%%%%%%%%%%

%%%%%%%%%%%%%%%%% BODY OF PAPER %%%%%%%%%%%%%%%%%%
\input{intro}
\input{mastar}

\input{colour_selection}
\input{results}

\input{classification}

\input{conclusion}

\section*{Acknowledgements}
% Entry for the table of contents, for this guide only
LH acknowledges support from the Science \& Technology Facilities Council (STFC) as well as the Data Intensive Science Centre in SEPnet (DISCnet). Numerical computations were done on the Sciama High Performance Compute (HPC) cluster which is supported by the ICG, SEPNet and the University of Portsmouth.

RY acknowledges support by the Hong Kong Global STEM Scholar scheme, the Hong Kong Jockey Club Charities Trust through the JC STEM Lab of Astronomical Instrumentation, the Direct Grant of CUHK Faculty of Science, and by the Research Grant Council of the Hong Kong Special Administrative Region, China (Projects No. 14302522: CUHK 14302522). 

J.G.F-T gratefully acknowledges the grant support provided by Proyecto Fondecyt Iniciaci\'on No. 11220340, and also from ANID Concurso de Fomento a la Vinculaci\'on Internacional para Instituciones de Investigaci\'on Regionales (Modalidad corta duraci\'on) Proyecto No. FOVI210020, and from the Joint Committee ESO-Government of Chile 2021 (ORP 023/2021), and from Becas Santander Movilidad Internacional Profesores 2022, Banco Santander Chile. 

Funding for the Sloan Digital Sky Survey IV has been provided by the Alfred P. Sloan Foundation, the U.S. Department of Energy Office of Science, and the Participating Institutions. SDSS-IV acknowledges support and resources from the Center for High Performance Computing at the University of Utah. The SDSS website is www.sdss.org.

SDSS-IV is managed by the Astrophysical Research Consortium for the Participating Institutions of the SDSS Collaboration including the Brazilian Participation Group, the Carnegie Institution for Science, Carnegie Mellon University, Center for Astrophysics | Harvard \& Smithsonian, the Chilean Participation Group, the French Participation Group, Instituto de Astrof\'isica de Canarias, The Johns Hopkins University, Kavli Institute for the Physics and Mathematics of the Universe (IPMU) / University of Tokyo, the Korean Participation Group, Lawrence Berkeley National Laboratory, Leibniz Institut f\"ur Astrophysik Potsdam (AIP),  Max-Planck-Institut f\"ur Astronomie (MPIA Heidelberg), Max-Planck-Institut f\"ur Astrophysik (MPA Garching), Max-Planck-Institut f\"ur Extraterrestrische Physik (MPE), National Astronomical Observatories of China, New Mexico State University, New York University, University of Notre Dame, Observat\'ario Nacional / MCTI, The Ohio State University, Pennsylvania State University, Shanghai Astronomical Observatory, United Kingdom Participation Group, Universidad Nacional Aut\'onoma de M\'exico, University of Arizona, University of Colorado Boulder, University of Oxford, University of Portsmouth, University of Utah, University of Virginia, University of Washington, University of Wisconsin, Vanderbilt University, and Yale University.

\section*{Data Availability}
Our results are available at the ICG MaStar page, \url{https://www.icg.port.ac.uk/mastar/}. 
MaStar spectra and accompanying data are available through SDSS DR17: \url{https://www.sdss.org/dr17/mastar/}.

\addcontentsline{toc}{section}{Acknowledgements}



\bibliographystyle{mnras}
\bibliography{C-MaStar} % if your bibtex file is called example.bib

\appendix
\input{appendix}

\end{document}

%% file: intro.tex
\section{Introduction}
\label{sec:intro}
Stellar populations in galaxies include stars across all evolutionary phases, which contribute to the integrated galactic spectrum according to their energetics, timescales of evolution and stellar spectral energy distribution. For calculating stellar population models \citep[e.g.][]{maraston_2005}, which aim at describing the integrated emission of complex stellar systems, a proper account of all stellar species is key. While energetics and timescales are usually known from stellar evolution calculations, stellar spectral energy distributions can either be taken from theoretical model atmosphere calculations \citep[e.g.][]{kurucz_1979} or from libraries of observed spectra of Milky Way stars \citep[e.g. MaStar, ][]{yan_etal_2019}. While both approaches are advantageous in different respects \citep[e.g. see extensive discussion in][]{maraston_and_stromback_2011}, the use of empirical spectra is unavoidable for certain stellar types. To this class belong Carbon-rich (hereafter C-rich) and Oxygen-rich (hereafter O-rich)-type stars, populating the short yet luminous Thermally-Pulsating Asymptotic Giant Branch (TP-AGB) phase of stellar evolution. The TP-AGB phase, with a maximum fuel consumption for $3~M_{\odot}$~stars~\citep{maraston_1998},was predicted and has been shown to be relevant for the spectral modelling of $\sim1$~Gyr populations, such as those featuring high-redshift galaxies and local star forming galaxies \citep{maraston_2005,maraston_etal_2006,capozzi_etal_2016,riffel_etal_2015,Liu_and_Luo_2023}. This is now confirmed spectroscopically by the detection of the strong spectral features typical of C-rich and O-rich TP-AGB stars, in the spectra of distant ($z\sim1-2$), massive ($M\sim10^{10}~M_{\odot}$) quiescent galaxies \citep[][]{lu_etal_2024}.

The challenges presented by TP-AGB C- and O-rich stars and their spectra are manyfold. Firstly, the whole TP-AGB phase only lasts $\sim 3$~Myr, meaning subphases tracing the stellar and envelope evolution along the phase are even shorter. The short timescales imply a paucity of calibrators for theoretical tracks and spectra. Next, C- and O-rich type spectra are featured by low temperatures, in general, emitting little flux below~$6000 \AA$\footnote{An exception are Carbon stars of secondary origin, see Sec. 4.3} and deep molecular absorptions, such as TiO, VO, CO, $\rm H_{2}O$, CN which are complicated to model. Finally, the TP-AGB phase is characterised by strong mass-loss and pulsation (on evolutionary and dynamical timescales), which adds further complications as the spectra are not static in most cases \citep{iben_and_renzini_1983, lancon_and_wood_2000, lancon_and_mouhcine_2002, rosenfield_etal_2016, hofner_and_olofsson_2018}. In summary, large stellar samples are needed to deal with C-O-rich type TP-AGB spectra.  

In this work we exploit the latest release of the MANGA Stellar Library MaStar \citep[MaStar][and Yan et al. {\it in prep.}]{yan_etal_2019} - the largest empirical stellar library assembled to date containing as many as 59,266 good quality spectra for 24,130 unique stars at a median resolution of R $\sim 1800$\footnote{These figures refer to the `good spectra' catalogue, but in our analysis we further include spectra not belonging to this cut.} - to hunt for the rare C- and O-rich type TP-AGB spectra. The wavelength range covered by MaStar - $3620 - 10350$ \AA\ - albeit still confined into the optical, is wide enough as to sample the unique features found in this type of spectra and allow a distinction between Carbon-rich and Oxygen-rich type stars. Below $1\mu$, the spectra of late, AGB O-rich stars are mostly featured by TiO (Titanium Oxide) and VO (Vanadium Oxide) bands (see discussion and Figure 4 in \citet[][]{lancon_and_wood_2000} and references therein). Carbon-rich spectra are quite different from the O-rich ones, being mostly featured, in the same spectral range below $1\mu$, by sharp CN (cyanide radical) bands. Broadly speaking, O-rich spectra display broader, more gentle absorptions, whereas C-spectra have harder, sharper features \citep[cfr. Figures 2 and 7 in ][]{lancon_and_mouhcine_2002}. Due to these difference, it maybe  possible, albeit time-consuming, to visually inspect all MaStar spectra for C- and O-rich type stars and perform a rough classification. On the other hand, we wish to obtain a more accurate, quantitative identification and classification and to define an automated procedure also in view of large scale surveys that will observe orders of magnitude more spectra than MaStar.

%The X-Shooter Spectral Library (3000 - 25,000 \AA, $7700 \leq$ R $\leq 11,000$) contains spectra for over 700 stars and a subset of 35 carbon stars \citep{chen_etal_2012}. In \citep{gonneau_etal_2015} the authors describe a technique to select 

In order to reach this goal, we explore broad-band colours based on the SDSS filters \textit{u, g, r, i, z} \citep{fukugita_etal_1996} to identify and separate C- and O-rich type spectra. While the use of only SDSS filters was previously explored in the literature \citep{margon_etal_2002, downes_etal_2004, green_2013}, the novelty here is that we adopt previously classified C- and O-rich type spectra \citep[from][ hereafter LM02]{lancon_and_mouhcine_2002} to calibrate the equation of our colour selection cut. This way we obtain a new and calibrated colour selection cut based on \textit{g$-$r} and \textit{g$-$i} to select C- and O-rich type spectra from any photometric survey\footnote{Different filters may require adjustment of the colour equation.}. We quantify the efficiency of our colour selection and of literature work \citep{margon_etal_2002}, in terms of purity and completeness, which demonstrate the increased accuracy we obtain.

We then perform spectral fitting of the colour-selected candidate MaStar spectra for C and O-rich types using our spectral fitting pipeline for MaStar spectra (Hill et al. 2022a) but now adopting as fitting templates empirical C-and O-rich spectra. With this procedure we could also identify a set of non-AGB, dwarf Carbon star spectra of secondary origin. 

It should be stressed that our classification metric is based on average spectral types for O-rich and C-rich stars (see LM02, Tables 1,2 for details on the component stars in each average), the latter mostly containing standard N-type AGB carbon stars. Hence by their use we can only probe this broad distinction and cannot attempt any finer spectral classification into (rare) Carbon sub-types such as J,R,S \citep[see][for a review on Carbon stars]{wallerstein_and_knapp_1998}. Note also that a proper analysis of these peculiar spectral types requires spectra with a higher spectral resolution than that of MaStar (Sec. 2) and a wavelength extension into the near-IR or a combination of both, i.e. spectral indices plus matched near-IR magnitudes as in the Carbon star classification for the LAMOST survey by \citet[][]{ji_etal_2016}, or using a combined GAIA-2MASS colour diagram as in \citet[][]{abia_etal_2020,abia_etal_2022}. In our work we make use of just SDSS optical colours and MaStar spectra.

The paper is organised as follows. In Section 2 we summarise the MaStar observations and describe our calculation of synthetic colours.
Section 3 presents our new colour equation for the selection of C- and O-rich stars.
Section 4 presents a quantitative evaluation of the performances of the new selection and of a previous selection for C-rich stars from the literature. Section 4 also reports the results of the identification.
Section 5 present a spectral classification of the candidates selected via photometry. Summary and conclusions are in Section 6.

%% file: mastar.tex
\section{MaStar, the MaNGA Stellar Library}
\label{sec:mastar}
\subsection{Generalities}
\label{sec:mastar-general}
MaStar (the MaNGA Stellar Library, \citep{yan_etal_2019} is currently the largest empirical stellar library of Milky Way stellar spectra, with extensive optical wavelength coverage and spectral types spanning the HR-diagram. Observations were carried out with the Baryon Oscillation Spectroscopic Survey (BOSS) spectrographs \citep{smee_etal_2013} mounted on the 2.5m Sloan Foundation Telescope \citep{gunn_etal_2006} at the Apache Point Observatory over a 6 year period. By observing in parallel to the APOGEE-2N survey \citep{majewski_16}, the MaStar survey was performed through the acquisition of optical spectra in the same field of view. By using fiber bundles, MaStar can provide spectra with a high median signal-to-noise of 96 per pixel. 
The resulting catalogue of stellar spectra contains 80,592 per-visit-spectra for 27,945 unique stars with the relatively wide wavelength coverage of $3620 - 10350$ \AA. Spectra are observed at a medium, wavelength dependent resolution (R) of 1800, which varies from fiber to fiber and from observation to observation \citep[see][for details on the resolution vector]{yan_etal_2019,law_21}. 

The unique observing strategy of MaStar has allowed for one of the most comprehensive stellar libraries in the literature, with an extensive coverage of the H-R diagram and rare combinations of atmospheric parameters \citep[][Chen et al. 2024, {\it submitted}]{chen2020,maraston_etal_2020, hill22a, hill22b, lazarz_etal_2022}. The full HR of MaStar can be seen in Figure~12 of \citet{hill22a}. To achieve this, stars with known stellar parameters from APOGEE (\citep{apogee}, LAMOST \citep{boeche_18}, SEGUE \citep{yanny_09}) were targeted to ensure a uniform coverage of atmospheric parameters and element abundance ratios. Then we complement this using photometry from Gaia DR1, Gaia DR2, Pan-STARRS1 \citep{chambers_16} and APASS to identify stars in rare and/or extreme parts of the stellar parameter space. Moreover, in early phases MaStar targeting was augmented by manual target selection for rare spectral types. This strategy allowed the sampling of stars in short-lived stellar phases, \textit{i.e.} the TP-AGB phase. It should be stressed that MaStar - by its observational strategy - is not meant to be a complete sample of spectra in any stellar phase. Relevant to this paper, MaStar observations were not performed by target selecting C and O-rich types specifically hence not tuned according to a star’s periodicity. Further details of the MaStar observing strategy and data reduction can be found in \citet{yan_etal_2019} and Yan et al. (in prep). The MaStar data products are described in section 6 of \citet{sdss_dr17} and are publicly available to download\footnote{\url{https://www.sdss.org/dr17/mastar/mastar-spectra/}}.

MaStar observations are split into two catalogues named `goodspec' and `badspec' \citep[see][]{yan_etal_2019}, containing 59,266 and 21,326 per visit spectra, respectively. The `goodspec' is the standard catalogue of good, ready-to-use spectra, while the `badspec' is a mixed bag of truly bad spectra due to e.g. missing pixels, but also unusual spectra, including emission-line spectra and rare types that were not obviously classified by the visual inspection, which was the main criterium to place spectra in either catalogue. Due to their peculiar spectra, C and O-rich type stars may exist in the `badspec' catalogue, which is why in our analysis we make use of both catalogues. However, in order to avoid using spectra with a large number of missing pixels, we remove spectra with a missing pixel fraction greater than $1\%$ from the analysis. This leaves 20,653 `badspec' spectra\footnote{Criteria for a spectrum to be labelled as 'bad' include: emission lines present in the spectrum, red upturn, poor flux calibration and missing pixels, unrecognised pattern \citep[][]{yan_etal_2019}.}. 

\subsection{Synthetic Magnitudes}
\label{sec:mastar-synthetic-photom}
In order to study a colour classification we need magnitudes for all our spectra and for the calibrating set by LM02, by XSL17 and for the theoretical Kurucz spectra. While available SDSS magnitudes could be used for the MaStar stars, they need to be calculated from the spectra for the other sets. Therefore it is more rigorous to re-calculate magnitudes for all spectra, MaStar and non MaStar ones. This way we make sure all values are homogeneously determined from the spectra with the same code, response functions and zero points).

MaStar observations are matched to the Gaia EDR3 survey\footnote{See section 6.4 in \citet{sdss_dr17} for a detailed description of how the Gaia photometry matching was performed.} and it turned out that 97 percent of MaStar spectra have corresponding Gaia data \citep{gaia_edr3}. Using the \citet{BJones_21} distances and 3D dust map values of \cite{green_etal_2019}, we correct spectra for galactic extinction before calculating the synthetic photometry. This is done using the \citet{fitzpatrick_1999} dust extinction law.

For each spectrum we then calculate the synthetic photometry in the AB system for the SDSS filters \textit{u,g, r, i, z} \citep{fukugita_etal_1996}. The synthetic photometry is calculated with the code of Maraston (2005), which performs a convolution integral of the spectrum with the response function normalised to the integral of the response function. We further adjust the filter transmission profiles of $u$~and $z$ to the MaStar wavelength extension (3630 -- 10,300 \AA\ ), which does not cover $u$~and $z$ entirely. We cut-off the filter curves of the standard \textit{u} band from 3040 \AA\ to 3622 \AA\ and of the standard \textit{z} band from 11600 \AA\ to 10300 \AA\, respectively. It should be noted that this does not affect our analysis because the same modified response function is applied to all spectra and also because the colour selection does not employ either affected bands $u$ or $z$. 

In addition to SDSS synthetic magnitudes, we calculate the $BVRI$ colours in the Johnson-Cousins system. In Section \ref{sec:linear_cut_colour}, we show how our selection of C-,O-rich spectra forms a unique sequence in the ($V-R$) vs. ($V-I$) colour-colour plane (Figure~\ref{fig:VR-VI}). Furthermore, we crossmatch all MaStar observations with 2MASS \citep{skrutskie_etal_2006} to obtain \textit{JHK$_{s}$} photometry. This returns 2MASS photometry for 99\% of the spectra we use. Details of how the crossmatch is performed can be found in \citet{sdss_dr17} and Yan et al. (in prep).

\begin{figure}
 \includegraphics[width=0.95\columnwidth]{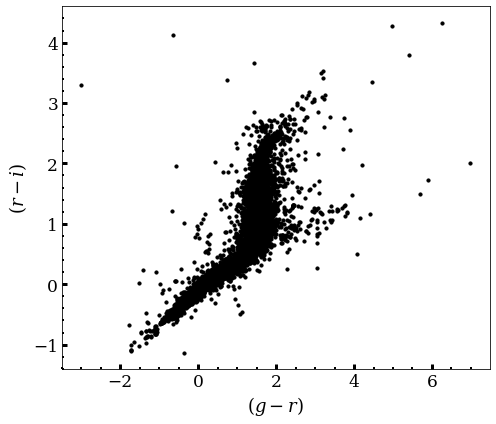}
 \caption{Distribution of all spectra used in this study shown as a colour-colour plot in $g-r$ vs. $r-i$~plane. Note the bifurcation at $g-r>2$.}
 \label{fig:gr_ri}
\end{figure}

In Figure \ref{fig:gr_ri} we show the distribution of all data, 
in a ($g-r$) vs. ($r-i$) diagram. We find stars as red as ($g-r$) = 6.5 and $r-i$) $\sim 4.5$ (for reference, in Gaia colours, MaStar contains stars as red as G$_{BP}$-G$_{RP} = 4.7$).
A clear bifurcation is seen at $g-r>2$, which it turned out to coincide with the C- and O-average colours, as we shall explain in Section \ref{sec:linear_cut_colour}.
To gain an overview of all five magnitudes, we plot their pairwise relationship in Figure \ref{fig:mags_pairplot}. The diagonal plots in grey show the distribution of each magnitude as a kernel density estimate. As expected, magnitudes that are closer in wavelength have a tight relationship and those at opposite ends of the spectrum, such as \textit{u} and \textit{z}, are less correlated.

\begin{figure*}
	\includegraphics[width=\textwidth]{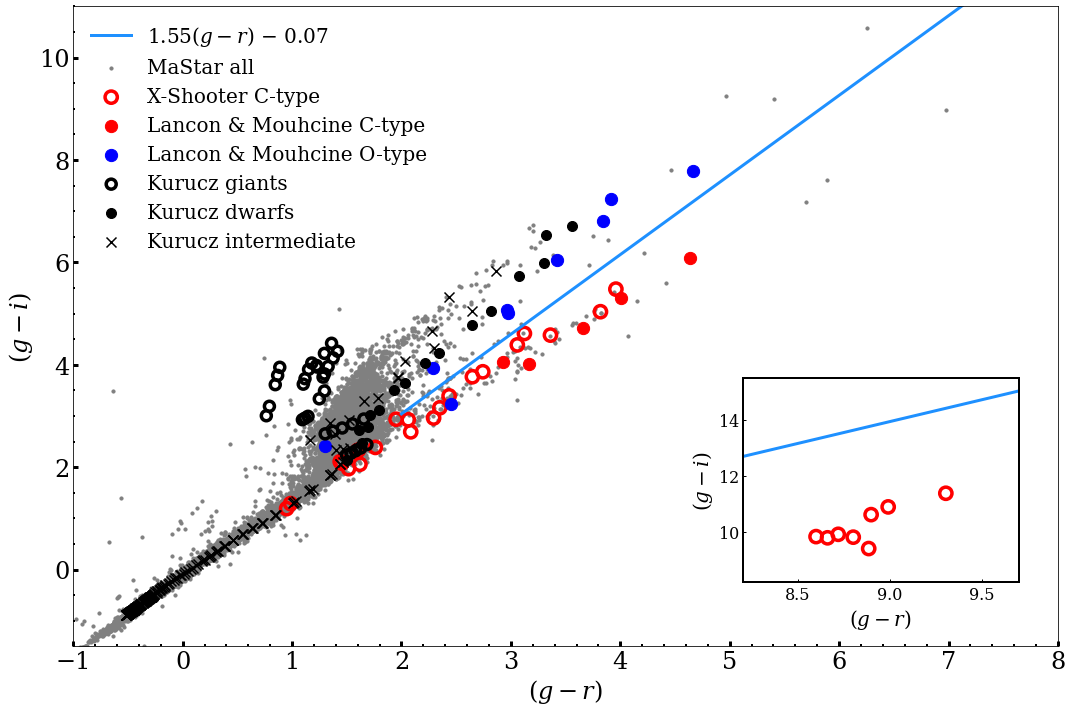}
	\caption{Colour selection for Carbon-rich and Oxygen-rich type stars overlaid to the whole colour-colour plane of MaStar spectra (grey points). Blue and red filled circles plot the 9 O-rich type and 5 C-type average spectra from \citet{lancon_and_mouhcine_2002}, red open circles are the colours of classified C spectra from the X-Shooter library \citep{gonneau_etal_2017}. The black markers are the calculated colours from \citet{kurucz_1979} model atmospheres, where giants (log g < 2.5) are open circles, dwarfs (log g > 4.7) are filled circles and intermediate types (2.5 < log g < 4.5) are crosses. The blue line represents our separation of C- and O-rich types, following the formula:  1.55($g-r$) -0.07. The C spectra colour sequence is unique, while some contamination from dwarfs is found in the O-rich type sequence. The lower right inset plot shows the location of the reddest C spectra from the X-Shooter Library.}
    \label{fig:colour-colour-literature}
\end{figure*}

%% file: colour_selection.tex
\section{Definition of a colour selection equation}
\label{sec:linear_cut_colour}
Here we study a selection in optical colour space able to identify and separate TP-AGB C and O stars and distinguish them from the rest of stellar species in virtue of their very red colours and peculiar spectra. As mentioned in the Introduction, a similar approach for the selection of C-stars is found in\citep[][]{margon_etal_2002}. In Section~\ref{sec:testing} we shall compare ours to this previous effort.

%
%\subsection{SDSS}
In order to obtain a colour selection cut able to identify C- and O-rich type TP-AGB spectra, we start by calculating the SDSS colours of known C- and O-rich type TP-AGB spectra. We use the \citet[][ hereafter LM02]{lancon_and_mouhcine_2002} C- and O-rich type spectra, which are average spectra for these stellar types in order to remove variability. These are binned into 5 and 9 bins for C-type and O-rich type respectively, with 1 the warmest and 5 the coldest for C-types and 9 for O-rich types. These average spectra are used to describe the TP-AGB phase in the stellar population models by \citet{maraston_2005}. We also use colours from the C-type spectra identified by \citet[][ hereafter XSL17]{gonneau_etal_2017} in the X-Shooter Library. 

After exploration of combinations of the five SDSS magnitudes, we conclude in favour of the ($g-r$) vs. ($g-i$) colour-colour plane. This combination allows for the identification of C- and O-rich type spectra and the separation between the two types best and minimise contamination from other spectral types. For the selection of C-type stars, we suggest a colour selection as follows:

% \begin{ceqn}
\begin{align}
    (g-r)> 2 \text{ and } (g-i) < 1.55(g-r) -0.07
    \label{eq:1}
\end{align}
% \end{ceqn}

For O-rich type stars we use a similar colour relation, where the inequality of the second equation is reversed:

% \begin{ceqn}
\begin{align}
    (g-r)> 2 \text{ and } (g-i) > 1.55(g-r) -0.07
    \label{eq:2}
\end{align}
% \end{ceqn}

%
The calibration is visualised in Figure \ref{fig:colour-colour-literature}, where blue and red filled circles are the 9 O-rich type and 5 C-type averaged spectra from LM02, red open circles are the colours of classified C spectra from XSL17 and grey points represent MaStar spectra. Also plotted as black symbols are colours of standard Kurucz spectra from atmosphere models \citep{kurucz_1979, lejeune_etal_1998}, further divided between giants (log g < 2.5), dwarfs (log g > 4.7) and intermediate types (2.5 < log g < 4.5)\footnote{We note that some Kurucz giants are offset from the main locus of stars. This does not affect our analysis.}. The blue line, described by Equation \ref{eq:1} and \ref{eq:2}, was determined by firstly fitting a linear model to each sequence of C- and O-rich types from LM02 and XSL17. The average value of the gradient of these two lines was then calculated to return the blue line as plotted \footnote{Note that there are several XSL17 spectra with ($g-r$)> 8, but we exclude these from the fit since they are beyond the colour range of MaStar. These are shown within the inset plot of Figure \ref{fig:colour-colour-literature}, demonstrating that the colour selection would still classify these as C stars.}. A clear separation is visible between C and O-rich type spectra, with the two sequences lying roughly parallel in this colour plane. Furthermore, we can see that the reddest MaStar spectrum corresponds to the coolest LM02 C-type 5 bin. The C-type colour sequence is unique, while some contamination from dwarfs is found in the O-rich type sequence. In Section \ref{sec:testing} we shall quantify the contamination in both sequences.

\citet{margon_etal_2002} developed a similar colour selection for C stars, based on their separation from the main locus of stars, firstly proposed in \citet{krisciunas_etal_1998}. The Margon et al. colour selection is: $15 <$ r $<19.5$ and (\textit{r$-$i}) $< -0.4 + 0.64$(\textit{g$-$r}). In this paper, we test their effectiveness for C star selection in MaStar (see Section 4.1). 

\begin{figure}
 \includegraphics[width=0.95\columnwidth]{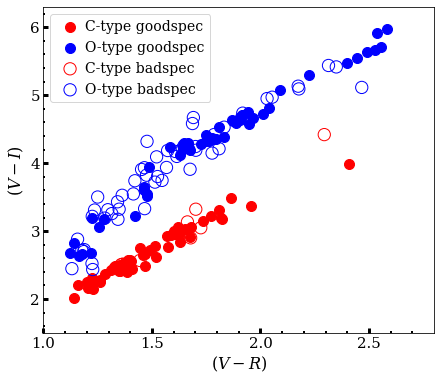}
 \caption{The MaStar sample of C- and O-rich type spectra plotted using the synthetic colours $V-R$~vs~$V-I$~in the Johnson-Cousins Vega system. C- and O-rich types are represented by red and blue colours, respectively, with data from the badspec sample plotted as open symbols.}
 \label{fig:VR-VI}
\end{figure}

 Figure~\ref{fig:colour-colour-literature} also allows us to address the question of how complete in C-rich and O-rich stars the MaStar sample is. The figure shows that MaStar data cover (in colour) all averaged types by LM02 and XSL17. Candidates in Mastar extend to even redder colours (grey points in Figure 2). Therefore we conclude that MaStar is able to cover the main broad spectral types of O-rich and C-rich stars as dedicated optical plus near-IR surveys such as e.g. the one by \textcolor{orange}{\citet[][]{lancon_and_wood_2000}} and also most part of the XSL17 C-stars. We shall return to this point in the Conclusions (Section~\ref{sec:summary}).
 
To conclude, the existence of unique colour sequences tracing C-rich and O-rich spectra is not a prerogative of SDSS colours. In Figure \ref{fig:VR-VI} we plot the selected MaStar C- and O-rich type spectra in a ($V-R$) vs. ($V-I$) colour diagram, which is the Johnson-Cousin analogue of Figure \ref{fig:colour-colour-literature}. Well-defined colour sequences trace and distinguish the colour loci of C-rich and of O-rich spectra. Both data from goodspec and from bad spec lie on these sequences (filled and open symbols, respectively). The better separation of the two types in $V-I$~with respect to $V-R$~arises because the I-filter (effective wavelength $\sim~8000$~Angstroms) is located around deep and broad TiO,VO absorption bands for O-rich stars, while C-stars, at the same wavelengths, are featured by narrower CN absorptions (cfr. Figure~\ref{fig:C-O-example-grid}). The $R$-filter (effective wavelength at $\sim~6400$~Angstroms) encompasses a relatively smooth region of the continuum for both types, which is why their colours are similar, but note the extension towards (V-R) colours for the later (i.e. coolest) O-rich types.

%% file: results.tex
\section{Results}
\label{sec:results}

\subsection{Quantitative assessment of different colour cuts.}
\label{sec:testing}
In this work we present a new colour equation for selecting C- and O-rich type stars. Our equation only uses SDSS photometric bands and is therefore cheap, not requiring any near-IR photometry nor spectra, and can be easily applied to a variety of galaxy surveys. In this section we quantitatively assess the power of our classification method plus the Margon et al. colour equation for C-types, using an homogeneous test sample.

To measure the effectiveness of the colour cut classifiers, we use the F1 and F-beta score metrics. The F1 score averages the precision and recall and is a useful metric when working with class labels that are not equally represented. In our case, there are only a few C- and O-rich type spectra compared to all other stellar types, making our class labels imbalanced. The F1-score metric is defined as:
\begin{ceqn}
\begin{align}
    \textrm{F1 } score = 2 * \frac{Precision * Recall}{Precision + Recall}
    \label{eq:6}
\end{align}
\end{ceqn}
where \textit{precision} measures the number of true positives as a fraction of all predictions and \textit{recall} is the number of true positives as a fraction of all data points in the target class, where the target class is labelled C- or O-rich type spectra\footnote{This is akin to the purity and completeness metrics used in astronomy.}. For this data, a true positive would be the case where a C- or O-rich type spectrum is predicted successfully by the classifier. 

Additionally, we consider the F-beta score, which allows one to adjust the importance of either precision or recall in the F1 score calculation. 
The F-beta score is defined as:
\begin{ceqn}
\begin{align}
\textrm{F-beta } score = \frac{(1+\beta^2)* Precision * Recall}{{\beta^2}*Precision + Recall}
    \label{eq:7}
\end{align}
\end{ceqn}
Since we are interested in how well the classifiers can select a specific class of stars out of all observations, we set beta to 0.5 which makes the output of the metric more sensitive to precision. The best value for the F1 and F-beta score is $1$ and the worst is $0$.

\begin{figure*}
 \includegraphics[width=1\textwidth]{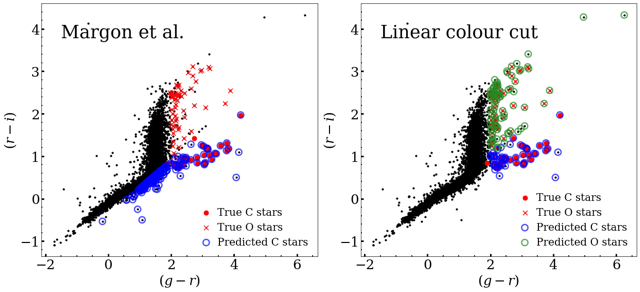}
 \caption{Comparison between colour selection methods. Left: the Margon et al. colour equation for C spectra only. Right: the colour equation presented in this paper for both C and O stars. The Margon et al. method returns an F1-score of 0.27, the ($g-r$) $> 2$, ($g-i$) < 1.55($g-r$) -0.07 cut returns 0.72 for C-types and 0.74 for O-rich types. Red data points show true C stars and blue circles are the predictions for each method. O-rich types are shown by red crosses and their predictions represented by green circles.}
 \label{fig:SVC-best-model}
\end{figure*}

\begin{figure*}
	\includegraphics[width=0.91\textwidth]{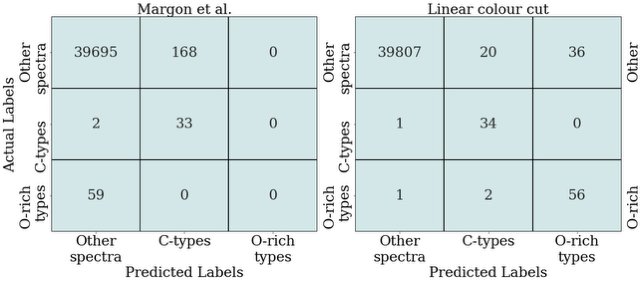}
	\caption{Confusion matrix showing how each method classified C-types, O-rich types and all other spectra.
 The values correspond to the results shown in Figure \ref{fig:SVC-best-model}.}
    \label{fig:confusion}
\end{figure*}

Figure \ref{fig:SVC-best-model} visualises the results of the testing, showing the ($g-r$) vs. ($r-i$) diagram for the two tested frameworks. 
%Note that this figure only shows half of the total data (39,957 spectra), since we are using a $50-50$ stratified split. 
%and the other $50\%$ was used to train the SVC model. Therefore, the test set represents  
The left-hand panel shows the results when using the colour selection for C-stars proposed by \citet{margon_etal_2002}, while the right-hand panel shows the C- and O-rich selection using the colour equations we propose (section 3, Figure 2).
%and the right panel shows the C- and O-rich selection from this paper using the SVC method (section 4). 
In each panel, the predicted C spectra are shown as blue open circles while red filled circles show the location of the 35 true C star spectra. In the right-hand panel, where we additionally test for O-rich spectra identification, we show the predicted O-rich type spectra as green open circles while red crosses represent the location of true O-rich type stars.

We now report and comment on the quantitative results for the iteration of data shown in Figure \ref{fig:SVC-best-model}. Starting from the left panel showing the Margon et al. method, the selection encompasses a large region of the main stellar locus that would be populated by non-AGB, dwarf C stars (see Section~\ref{sec:C-dwarf}). Nevertheless, 33 of the 35 true C stars are selected, resulting in a recall score of 0.94, but low precision of 0.16. Our newly proposed colour cut (right-hand panel) is able to select all but one C star, returning a recall score of 0.97 and precision of 0.61 due to 20 false positives. It can also identify O-rich type stars in the region $2 <$ ($r-i$) $< 3$ and miscellaneous stellar types in between with a recall of 0.95 and precision of 0.61. 

The confusion matrices for the data represented in Figure \ref{fig:SVC-best-model} are shown in Figure \ref{fig:confusion}.

In Table \ref{table:identification_stats} we show the mean F1 and F-beta scores. 
Focusing on C-star methods as this is covered by both approaches, the Margon et al.'s method returns a mean F1-score of 0.27 while our cut of ($g-r$) $> 2$, ($g-i$) < 1.55($g-r$) $-0.07$ returns 0.72.  
This trend is repeated for the mean F-beta scores. Therefore the method presented in this paper is an improvement over the Margon et al. method. However, as mentioned, their method does capture the colour region where non-AGB C stars exist; this may be useful for general C star identification, but not for our analysis. For O-rich type stars, the calibrated colour equation method we put forward returns a mean F1-score of 0.74.

\begin{table}
\caption{Score metrics for the colour selections. For both methods we report the star type, the mean F1-score, the mean F1-score and the variance. We also show the results of the colour cut used to identify O-rich type stars. The column `Type' signals whether we are considering C- or O-rich type spectra.} % title of Table
\centering % used for centering table
\begin{tabular}{c c c c } % centered columns (4 columns)
\hline\hline %inserts double horizontal lines
Model & Type & Mean F1 & Mean F-beta \\ [0.5ex]
%heading
\hline\hline % inserts single horizontal line
Margon et al. & C & 0.27 & 0.19 \\ \hline% inserting body of the table
($g-r$) $> 2$ and & C & 0.72 & 0.62  \\ 
($g-i$) < 1.55($g-r$) $-0.07$ & &   \\ \hline
($g-r$) $> 2$ and & O & 0.74 & 0.66 \\
($g-i$) > 1.55($g-r$) $-0.07$ & &   \\ \hline
\hline %inserts single line
\end{tabular}
\label{table:identification_stats} % is used to refer this table in the text
\end{table}

Finally, a confusion matrix for each classifying method is provided in Figure~\ref{fig:confusion}.
The values correspond to the results shown in Figure \ref{fig:SVC-best-model}.
\subsection{Identification.}
\label{sec:identification}
Using our newly defined colour equation we have identified C-rich and O-rich type candidate spectra in MaStar. Identifications have been confirmed with SIMBAD\footnote{https://simbad.u-strasbg.fr/simbad/sim-fbasic} and with visual inspection.
In detail, we have identified 41 TP-AGB C-stars represented by 69 spectra observations and 87 O-rich type stars, represented by 118 spectra observations. Of these 36 C-rich stars and 38 O-rich stars have a SIMBAD classification. 

\begin{figure*}
 \includegraphics[width=0.95\textwidth]{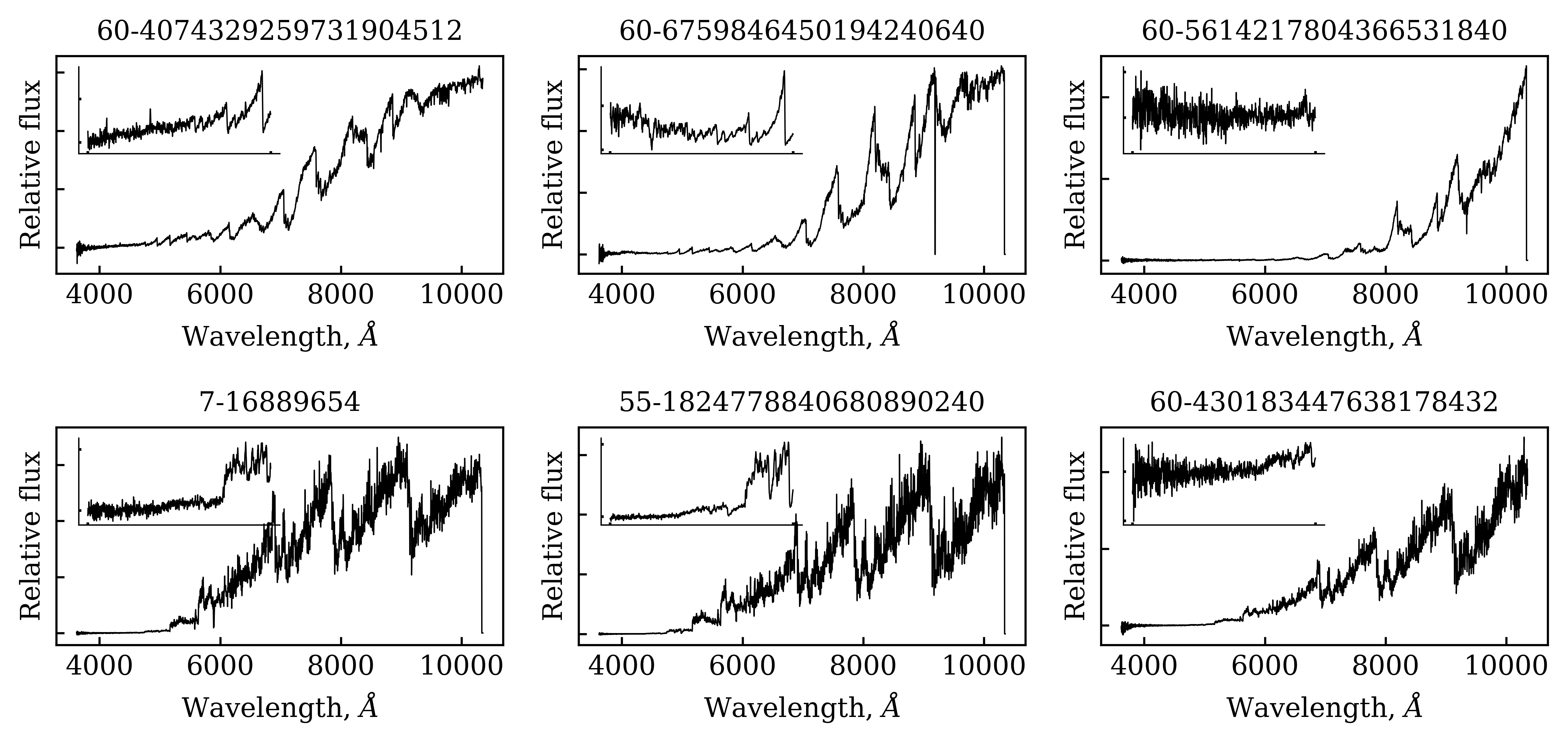}
 \caption{\textit{Example O-rich (top) and C-type (bottom) spectra identified in this work. Within each subplot we show a closer view of the spectrum in the wavelength range $4000 - 5000$\AA. The title of each subplot shows the MaNGA ID for the corresponding spectrum.}}
 \label{fig:C-O-example-grid}
\end{figure*}

In Figure \ref{fig:C-O-example-grid} we show three examples of O-rich spectra in the top row and three examples of C-types in the bottom row. The O-rich types in the top row display the strong TiO bands at 7200 \AA, 7700 \AA\ etc. and VO bands at 7400 \AA\ and show a gradual decrease in  T$_{\rm eff}$ from left to right. The defining spectral features of C-type spectra are the sharp CN bands and the absence of molecular bands such as H$_{2}$O and TiO. As above, the spectra show a decrease in T$_{\rm eff}$ from left to right. Furthermore, within each subplot we show a zoom in of the spectrum in the wavelength range $4000 - 5000$\AA. The C- and O-rich spectra for all stars is shown in Figures \ref{fig:C-star_spectra-all} and \ref{fig:O-star_spectra-all}, respectively. Spectra from the goodspec file are in black and those from the badspec file are blue. In Table \ref{table:params-all}, a sample of C- and O-rich spectra are represented, where the row number matches the order that the spectra are plotted, going from left to right, top to bottom. The full table can be downloaded at \url{http://www.icg.port.ac.uk/mastar/}. In the same table we provide the MaStar identification, median signal-to-noise ratio per pixel, star coordinates in degrees, calculated colours in both SDSS and Johnson-Cousins systems, whether the spectrum is from the badspec catalogue (flag=1) or not (flag=0), the C- or O-rich classification and the class given by full-spectral fitting with LM02 or XSL17 templates (see Section 5).

%% file: classification.tex
\section{Spectral Classification into C and O-rich subtypes.}
So far we have identified a set of candidates C-rich and O-rich spectra by only using photometry, then confirmed their nature via visual classification and quantitative metrics (plus existing tools like SIMBAD). Here we shall perform a spectral classification of the spectra of these candidates selected via photometry.

The distinction between C-rich and O-rich type spectra is relatively straightforward and differentiation can be done by eye using the prominent bands of VO, C$_{2}$, CN and TiO. However, to gain an estimate of atmospheric properties such as temperature, it is necessary to use the spectral energy distribution through spectral fitting or by photometry matching. For the cold C- and O-rich type spectra, the spectral features that are most sensitive to changes in effective temperature and chemical abundance are located in the near-IR. For example, the C$_{2}$H$_{2}$ and C$_{3}$ features at $3\mu$m and $5.1\mu$m are found to be most sensitive to T$_{\rm eff}$ \citep{paladini_etal_2011} and the C/O ratio \citep{Jorgensen_etal_2000}. For this reason, in our paper we use spectra that have been classified into temperature bins based on their near-IR photometry as templates for spectral fitting in the wavelength range of MaStar spectra.
\subsection{Empirical Templates}
\label{sec:empirical-templates]}
In order to classify the identified spectra, we look to comparing our observations with other classified empirical spectra. We compare with two libraries of lower (R $\sim1100$) and somewhat higher (R $\sim2000$) spectral resolution that brackets the MaStar median resolution of R $\sim 1800$. The lower resolution library is provided by LM02, whose colours we have used in Section 3, who - we recall - provide averaged spectra of C- and O-rich type stars from the Milky Way and Magellanic Clouds. We focus on this library in particular as they have been used to supplement the Maraston models since first introduced in \citet{maraston_2005} and in their models since. The averaged spectra are derived using individual stellar spectra as described in \citet{lancon_and_wood_2000}, at a resolution of R $\sim 1100$ and wavelength range $0.5 - 2.4$ $\mu$m. The motivation for averaged spectra is due to uncertainties in individual stars from thermal pulses, stellar wind and mass loss. LM02 sort 63 individual spectra of O-rich type stars into 9 bins of 7 spectra each, with bin number increasing with decreasing effective temperature, from $T_{\rm eff}=3930~K$ for the first bin O1 to $T_{\rm eff}=2430~K$~for the last bin O9 (Table 3 of LM02). These bins are sorted based on the broad band colours (I - K) which they find to correlate well with effective temperature (see Figure 11 of \citet{lancon_and_mouhcine_2002}). For C-type stars, 5 bins are used, where bins 1 - 3 contain 6 spectra each. Bin 1 includes spectra for one S/C-type star. Bin 4 contains the spectrum of R Lep at maximum light\footnote{The optical part of the spectrum is an average of bins 3 and 5.} and bin 5 the average of 3 spectra from R Lep at minimum light. They sort these based on the broad band colours (R - H) and C/O abundance ratios. Unfortunately we do not have such abundance ratio information for MaStar spectra, so a similar binning approach is not possible. The correspondence to temperature for the C-bins is $T_{\rm eff}=3200~K$ for the first bin C1 to $T_{\rm eff}=2000~K$~for the last bin C5 (Table 3 of LM02).

The second empirical library used in the classification is provided by the X-Shooter Spectral Library \citep{gonneau_etal_2017} (hereafter XSL17). They provide spectra of only C-type stars at a resolution R $\sim 2000$ and wavelength range $0.4 - 2.4$ $\mu$m. To estimate atmospheric properties, XSL17 fit their observations with theoretical spectra based on the models of \citet{aringer_etal_2009} that were calculated for their analysis. Their C-type stars are binned into four groups based on broad band colours: the bluest in group 1 (J - K$_{s}$) $< 1.2$, group 2 $1.2 <$ (J - K$_{s}$) $< 1.6$ and the reddest in groups 3 and 4 with (J - K$_{s}$) $> 1.6$. The distinction between the reddest groups is whether the $1.53$ $\mu$m absorption feature is present. Without this feature, stars are placed in group 3 and all other red stars place in group 4. As our wavelength range does not cover this feature we aggregate these groups and label all templates with (J - K$_{s}$) $> 1.6$ as group 3.

\subsection{Spectral Fitting Method}
\label{sec:spectral-fitting}
In this paper we adapt our spectral template fitting method which we developed for MaStar spectra and which we recall below, to work with the empirical templates described in Section~\ref{sec:empirical-templates]}.
Our method to determine stellar parameters - $logg, T_{\rm eff}, [Z/H]$ - for the observed MaStar spectra is based on a single stellar template fitting approach over the whole wavelength extent of the MaStar spectra, using as fitting templates theoretical stellar spectra from model atmospheres. This method, first described in \citet{maraston_etal_2020} was later developed to incorporate MCMC sampling in \citet{hill22a} and also expanded to add the estimation of the $[\alpha/Fe]$~parameter alongside the fundamental stellar parameters quoted above\citep{hill22b}. 
The full spectral fitting is performed using the penalized pixel-fitting method (pPXF, \citet{cappellari_and_emsellem_2004, ppxf}) to fit templates to data. This algorithm performs a $\chi^{2}$ minimisation by comparing the observed spectra with the templates. The minimum $\chi^{2}$ is found by solving the quadratic programming problem described by equation 20 in \citet{ppxf}. Furthermore, the algorithm adds a penalisation term as the line of sight velocity distribution (LOSVD) deviates from a Gaussian shape. This reduces the noise in recovered kinematics when fitting for stellar populations and is the original functionality of the algorithm. However, in the case of single stars, the parameterisation of the LOSVD can be used to correct for velocity offsets in the observations. Furthermore, by using multiplicative polynomials we can normalise the continuum and account for inaccuracies in spectral calibration caused by dust. This is particularly import for the AGB stars that can have dusty envelopes.

For the estimation of stellar parameters in this article we adopt the single template approach due to the scarcity of templates and the discreteness of their parameters. As fitting templates we adopt the two sets described in Section~\ref{sec:empirical-templates]}, namely the LM02 empirical binned C-rich and O-rich spectra and the XSL17 empirical binned C-rich spectra. Therefore it is important to stress that with this fitting we do not determine explicit stellar parameters, rather we determine the best fitting bin, which in turn correlates with temperature via near-IR colours as discussed in the papers where the binning is proposed, namely \citet{lancon_and_mouhcine_2002} and \citet{gonneau_etal_2017}. 

After correcting the observed spectra for Milky Way reddening (as described in Section~\ref{sec:mastar-synthetic-photom}) we treat each of them as a single star, assuming that no binaries exist in either templates or observations. For the comparison with LM02 templates, MaStar spectra are firstly convolved to match the template resolution of R $\sim1100$. When comparing to XSL17, the resolution is close enough to MaStar's such that we do not require this step. We then minimise the $\chi^{2}$ between each template and the observation. By selecting the template with the lowest $\chi^{2}$ we can determine the spectral type as assigned by either LM02 or XSL17. We stress that the two sets of templates are used for independent $\chi^{2}$ fitting.
In Section~\ref{sec:results-spectrafit} we shall show and discuss the results of our spectral fitting analysis.

\subsection{Results of Spectral Fitting}
\label{sec:results-spectrafit}
\begin{figure*}
	\includegraphics[width=0.95\textwidth]{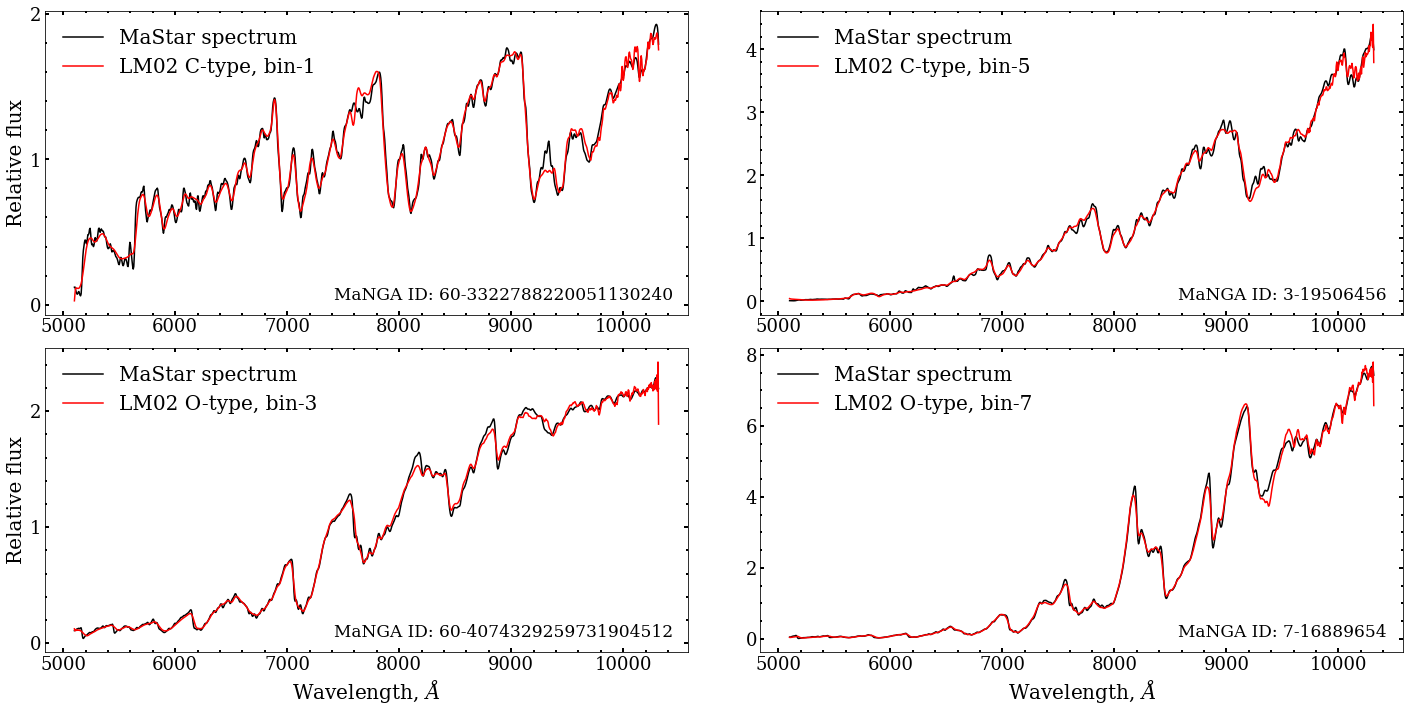}
	\caption{\textit{Top row}: Example full spectral fits of C-type MaStar spectra using the LM02 averaged spectra as templates. MaStar spectra have been convolved to match the spectral resolution of LM02 (R $\sim1100$). Shown are fits of the warmer bin-1 and cooler bin-5 spectra. \textit{Bottom row}: Same as above but for bin-3 and bin-7 O-rich type spectra.}
    \label{fig:LM02_fits}
\end{figure*}

\begin{figure*}
	\includegraphics[width=0.95\textwidth]{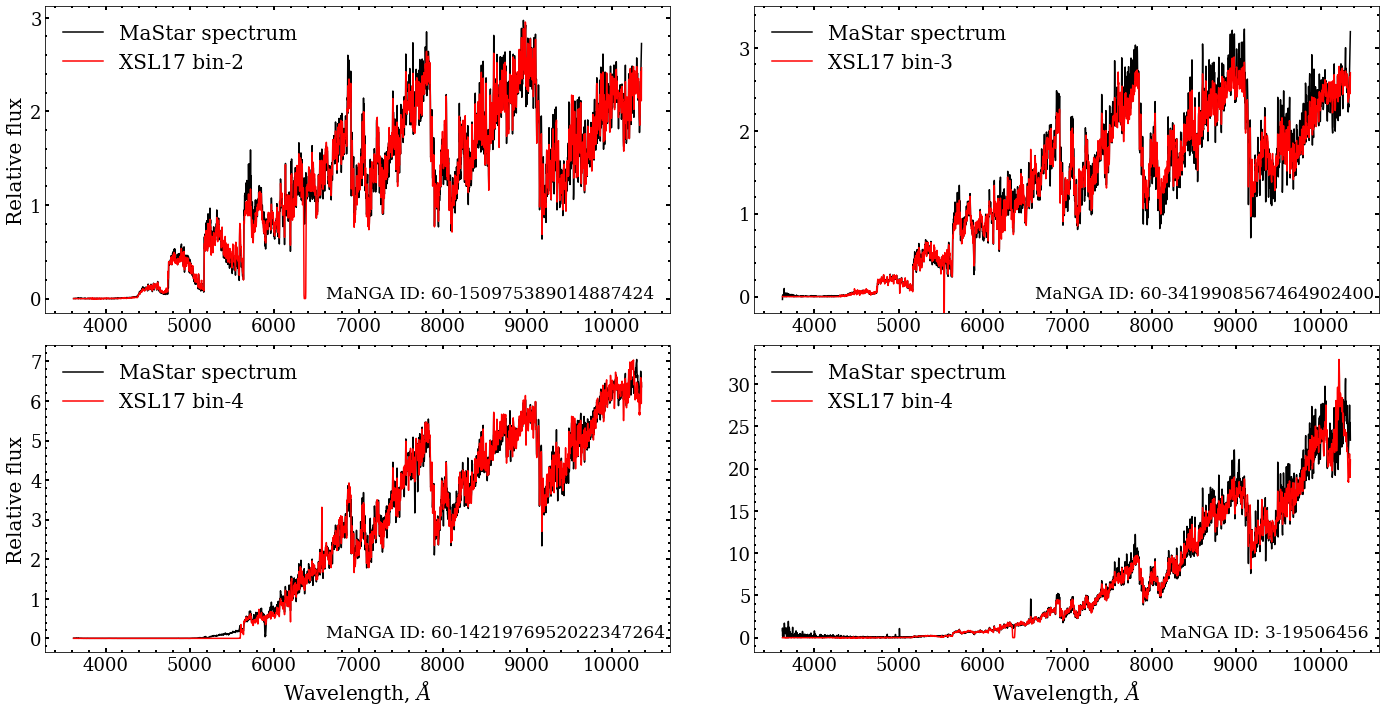}
	\caption{Example model fits of C-type spectra using the XSL17 spectra as templates, for the four temperature bins defined in \citet{gonneau_etal_2017}.}
    \label{fig:XSL17_fits}
\end{figure*}

In Figure \ref{fig:LM02_fits} we show example spectral fits of C (top row) and O (bottom row) stars using the LM02 averaged spectra as templates. In the top left is an example of the warmest bin in LM02 fit to a MaStar C star spectrum and the top right shows one of the coolest C spectra in our library, fit with LM02 bin-5. Below are example fits using the LM02 O type spectra as templates, the warmer bin-3 and cooler bin-7 are shown. Not all O type spectral fits were successful due to noise and our limited template bank. We therefore omit these poor fits based on their $\chi^{2}$ from the final catalogue. In Figure \ref{fig:XSL17_fits} we show example fits from fitting C star spectra to the XSL17 data. In general, the XSL17 spectra work well as templates and the assigned bin corresponds with the SED shape of each spectrum. A sample of C- and O-rich type MaStar IDs and their corresponding template fit from either LM02 or XSL17 are in Table B1.

\begin{figure}
	\includegraphics[width=1\columnwidth]{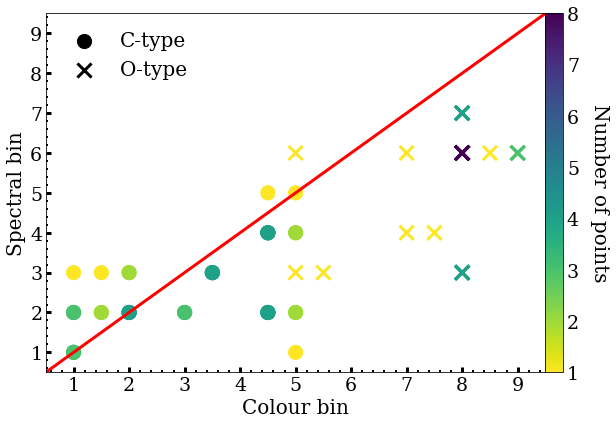}
	\caption{A comparison of the selected bin type when using colours to determine the spectral type of C- and O-rich types compared to our spectral fitting method. The thresholds for the colour bins are taken from LM02, corresponding to cuts in ($R-H$) for C-types and ($I-K$) for O-rich types. The colour of each data point shows where spectra overlap and the red diagonal line represents a perfect correlation.}
    \label{fig:colour-vs-specfit}
\end{figure}

Using the synthetic photometry in the Johnson-Cousins system (BVRI) and the 2MASS \textit{JHK$_{s}$} photometry, we are able to assign each spectrum according to the bins defined in LM02 for C- and O-rich types. Specifically, C-types are categorised based on their ($R-H$) colour and O-rich types using ($I-K$). In Figure \ref{fig:colour-vs-specfit}, we compare this bin number based on colour to the result of the spectral fitting approach previously described. This shows some correlation between the two methods. This result is encouraging as it shows that the spectral fitting method using the binned templates is accurate to within a few bins for most spectra and can be used alternatively to near-IR colours when these are not available. We note a preference towards later bins, i.e. colder temperatures using the colour binning. This maybe due to the fact that the spectral fitting is performed on the MaStar wavelength range, which - by being limited to $10,300~\AA$ - fails to assess the actual spectral type for the coldest spectra whose energy emission peaks further from the maximum wavelength covered by MaStar. It will be interesting to probe this speculation when near-IR spectra for our sample will be available.

\begin{figure*}
	\includegraphics[width=0.95\textwidth]{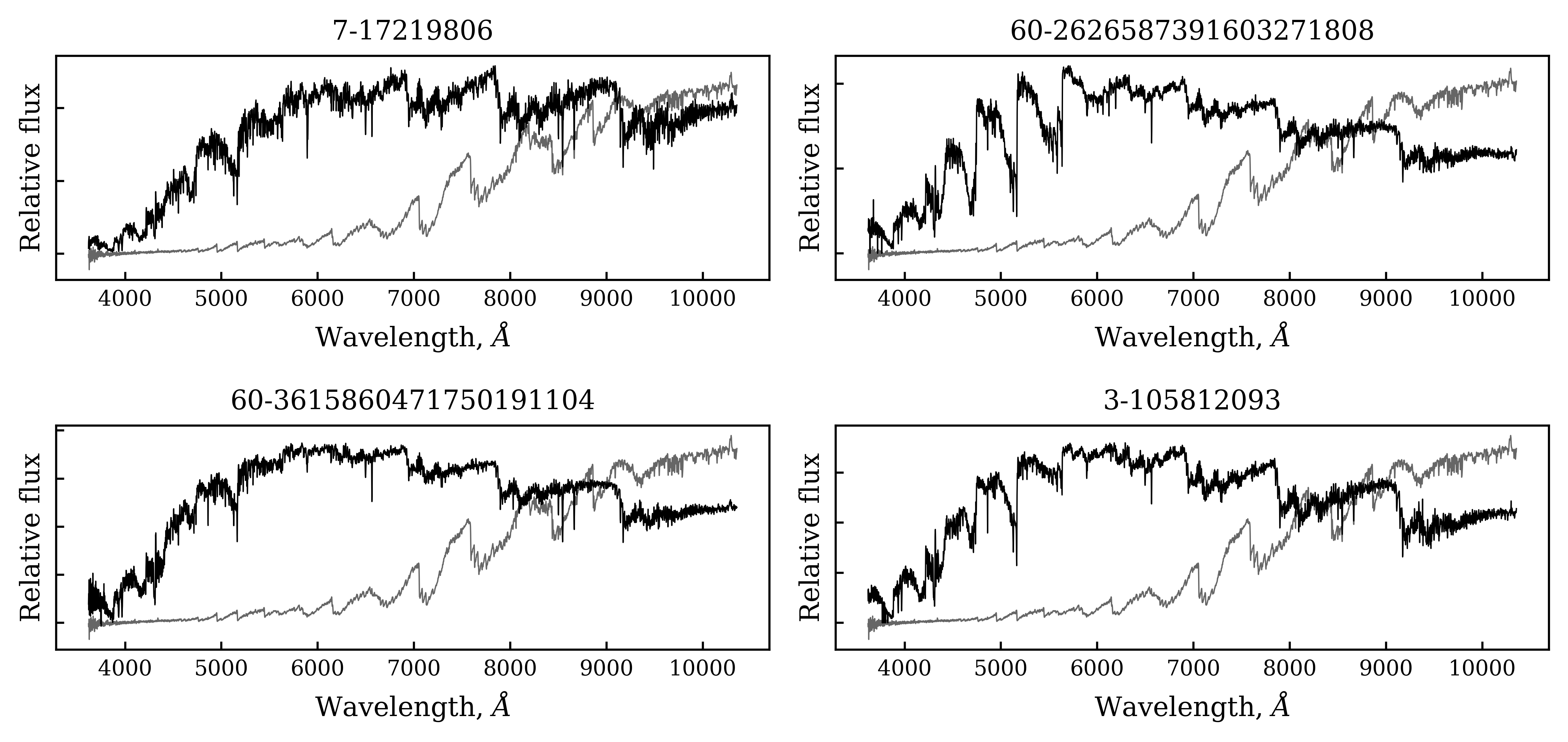}
 	\caption{Dwarf C-type spectra (black) in MaStar that are detected through our methods, but are excluded in the statistics since our focus is on AGB stars. The spectrum for an AGB C-type star (MaNGA ID 60-4074329259731904512) is shown in grey as a comparison in each panel. Note that the magnitude of the flux is not equivalent between the AGB and dwarf types and has been adjusted for the comparison. The title of each subplot shows the MaNGA ID for the corresponding dwarf C-type spectrum.}
    \label{fig:Dwarf-C-star_spectra}
\end{figure*}

\subsection{Effect of spectral variability.}
\label{sec:variability}

In our analysis we consider each individual spectrum because we aim at classifying each spectrum collected by MaStar. This means we consider different spectra for the same star. The aim at analysing average spectra for a same star could be pursued in the future. 

Here we report our findings regarding spectral variability. Of the 69 C stars and 118 O stars we have identified, there are 88 stars that have multiple observations. We do not find any instance where magnitudes are identical. 

We checked if and how the spectral classification changes between multiple observations. With respect to the X-shooter templates, we find 2 stars which have a variable classification between observations. The uncertainty is one classification interval in the XSL classification scheme. With respect to the Lancon \& Wood templates, we find 5 stars that vary between observations. These have larger variations between observations, which could be due to the lower resolution of the L\&W templates or to intrinsic variations.

In total, we find 7/88 stars whose spectral classification depends on the spectrum visit adopted. This is less than $10\%$, which supports our suggestion that stellar variability is not much of an issue in the optical. Detailed stellar IDs and classification bins are reported below.

XSL:
mangaid:60-3405514654587957760, classification:[2 3 3]\\
mangaid:60-3393303371074313216, classification:[3 2 3]\\

L\&W:
mangaid:55-1824778840680890240, classification:[5 1 2]\\
mangaid:55-2191637904670592896, classification:[1 4]\\
mangaid:55-2179423086398277760, classification:[3 1 1]\\
mangaid:60-6759846450194240640, classification:[4 6]\\
mangaid:60-430146270400578688, classification:[4 4 5]\\

\subsection{Detection of non-AGB, dwarf Carbon-type spectra}
\label{sec:C-dwarf}
During our spectral analysis we discovered a number of spectra with Carbon features, but a higher temperature as qualitatively deduced from their spectra. These are "non-AGB" or dwarf carbon stars, they sit in the main sequence and are thought to form from mass transfer in a binary system. In the formation scenario for classical, AGB Carbon stars \citep{iben_74, iben_and_renzini_1983}, the Carbon is pushed into the stellar atmosphere from successive dredge up events, in particular the third dredge up, a recurring event which moves carbon from the inner layers to the atmosphere thereby increasing the Carbon abundance relative to oxygen in the stellar atmosphere. The formation process of a dwarf, non-AGB C star is instead thought to be a consequence of carbon enhanced material from a classical TP-AGB carbon star being accreted by the dwarf companion. The TP-AGB star eventually evolves to a white dwarf via mass loss, remaining undetectable at optical wavelengths, leaving a Main Sequence star with Carbon features \citep{dahn_77, mcclure_90,roulston_22}.

Figure \ref{fig:Dwarf-C-star_spectra} shows that the spectra of dwarf C-types (black) display similar carbon bands to what is observed in AGB C-types (grey), but with a warmer continuum due to the higher effective temperature.

Non-AGB C-type spectra are not pertinent to AGB modelling (and cannot be fitted meaningfully by AGB C-type templates as we do here, due to a vastly different continuum). 

Therefore we do not include them in our analysis. On the other hand, these are unusual spectra that are interesting in themselves and for other astrophysics scopes. The IDs of the detected spectra are provided in Figure \ref{fig:Dwarf-C-star_spectra}.

%% file: conclusion.tex
\section{Summary and Discussion}
\label{sec:summary}
We have exploited the vastness of the SDSS-IV/MaNGA Stellar Library, MaStar ($\sim60,000$ spectra, covering the wavelength range $3620 - 10350$ \AA) in order to hunt for rare C- and O-rich type TP-AGB stars to facilitate the future modelling of intermediate age ($\sim1$ Gyr) stellar population. These stars feature unique spectral patterns, which help break well-known degeneracies in the interpretation of integrated galaxy spectra, such as the age/metallicity and age/dust degeneracies (e.g. \citet{maraston_2005}, Maraston et al. 2006). This aids the interpretation of extra-galactic stellar population spectra. Strikingly, spectral features typical of C-rich and O-rich stars evolving on the TP-AGB have recently been detected on the spectra of distant ($z\sim1-2$), massive ($10^{10}~M_{\odot}$), quiescent galaxies (Lu et al. 2024). Galaxy ages determined by fitting stellar population models including the contribution from the TP-AGB phase are indeed around 1 Gyr and the galaxy spectra cannot be fitted satisfactorily without a TP-AGB contribution, by e.g. changing the model metallicity, age, dust or star formation history. This fulfills the promise that the TP-AGB phase helps breaking degeneracies in the interpretation of the integrated spectra of stellar systems.

While the low-temperatures typical of TP-AGB stars ($T_{\rm eff}<4000$\;K) are best probed with near-IR surveys capturing the peak of their energy emission and most prominent spectral features (e.g. \citet{lancon_and_wood_2000}), MaStar covering spectra from 6000 to 10300 \AA\ already capture CN, TiO and VO bands and - as we show here - allows the identification of C,O type spectra just using optical SDSS colours. 
In addition, the large number of spectra probed by MaStar allows us to beat the expected low-number statistics for TP-AGB stars due to the short timescales involved ($\sim3$~Myr). Our results will allow a cheap and quick identification of these exotic stellar spectra and the exploitation of optical surveys without the need to obtain extra (and usually expensive) near-IR data. 

Our method seeks to identify the colour space occupied by these stars in SDSS colours in MaStar and we started by calculating the expected colours of named C and O stars from previously published catalogues. In this way we could robustly identify their colour loci and found that they stick out from the rest of spectra in a ($g-r$)~vs.~($g-i$)~colour displaying a clear bifurcation, which further allows a neat separation between the two types. Our newly proposed cuts are: ($g-r$) > 2 and ($g-i$) < 1.55($g-r$) - 0.07 for C-rich types and ($g-r$) > 2 and ($g-i$) > 1.55($g-r$) - 0.07 for O-rich types.
When compared with previous results for C stars \citep{margon_etal_2002}, we find our cut to be more effective. This is probably the result of our calibration of the colour selection using the colours of known C- and O-type spectra. Furthermore, new with respect to these past efforts, it also allows the identification of O-rich type spectra, albeit with more contamination. 

With our colour-colour cut ($g-r$) > 2, ($g-i$) > 1.55($g-r$) $-0.07$ we have identified 41 AGB C-type stars, which are represented by 69 spectra as some stars are observed multiple times and 87 O-rich type stars, for which MaStar contains 118 spectra. Of the C and O stars identified here, 5 C stars and 49 O stars do not have a SIMBAD classification and are therefore new identifications from our analysis, which would be interesting to confirm with other observational facilities in the future.

Further, we put forward a new approach to spectral classification for C-,O-rich spectra. This is based on the same full spectral fitting approach we developed for determining the atmospheric stellar parameters for MaStar \citep{hill22a}, but in this case we use as fitting templates the empirical C,O spectra of \citet{lancon_and_mouhcine_2002} and \citet{gonneau_etal_2017}. This way we are able to assign a spectral class to each C- and O-rich type MaStar candidate, with correlation with effective temperature. Using the colour bins to define the spectral type, as defined in \citet{lancon_and_mouhcine_2002}, we are able to compare to the spectral fitting method. We find some correlation between the two methods and that there is a preference to colder temperatures when using the colours to determine spectral type.

Our work focuses on optical spectra which is what is available in MaStar. On the other hand, galaxies emit most energy at optical/near-IR wavelength, therefore for studying galaxy evolution it is important to include the contribution of stellar phases at these wavelengths. As well known, TP-AGB stars emit most of their light in the near-IR and the coldest and dustier among them, which correspond to the shortest lived phases, at even longer wavelength \citep[e.g.][]{dellagli_etal_2017}. In this sense our work sits besides long wavelength surveys explicitly dedicated to the study of AGB stars, e.g. the Nearby Evolved Star Survey (NESS\footnote{http://evolvedstars.space}), a (sub-)mm, multi-band, volume-limited survey of mass losing AGB stars. It will be interesting in the future to crossmatch our detections with other catalogues of C- and O-rich spectra such as those mentioned in the Introduction \citep[][]{ji_etal_2016,abia_etal_2020,abia_etal_2022} and with NESS in order to reconstruct the multi-wavelength spectrum of TP-AGB stars. 

The C-,O-rich TP-AGB spectra that we have identified here will enable a more accurate definition of young and intermediate-age MaStar-based stellar population model spectra (Maraston et al. (in prep)). In addition, the colour cut we developed can be applied to other stellar and galaxy/cosmological surveys employing the same, or similar, filters to either detect even more C,O type spectra or study stellar contamination for high-redshift galaxy surveys.

%arXiv:2403.07414v1

%% file: appendix.tex
\section{Magnitude Comparison}
\label{sec:mags_pairwise}

\begin{figure*}
	\includegraphics[width=0.91\textwidth]{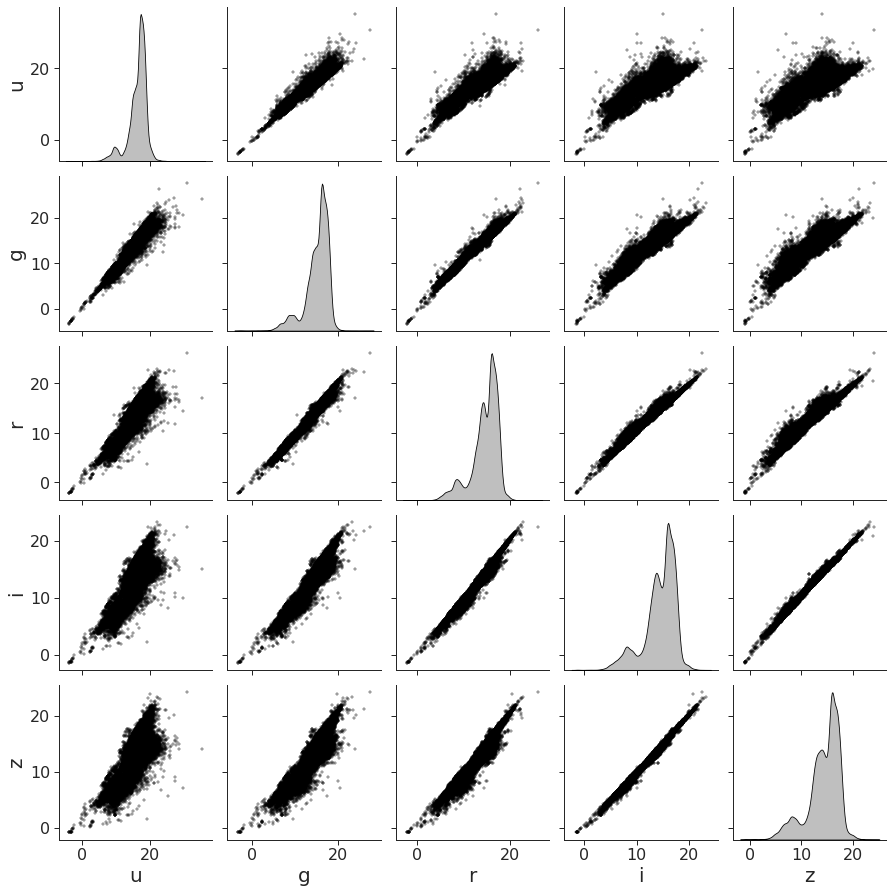}
	\caption{Pairwise comparison of the five magnitudes used for the C-type and O-rich star classification.}
    \label{fig:mags_pairplot}
\end{figure*}

\section{C and O star Spectra}
\label{sec:appendix-cstars}

\begin{figure*}
	\includegraphics[width=0.91\textwidth]{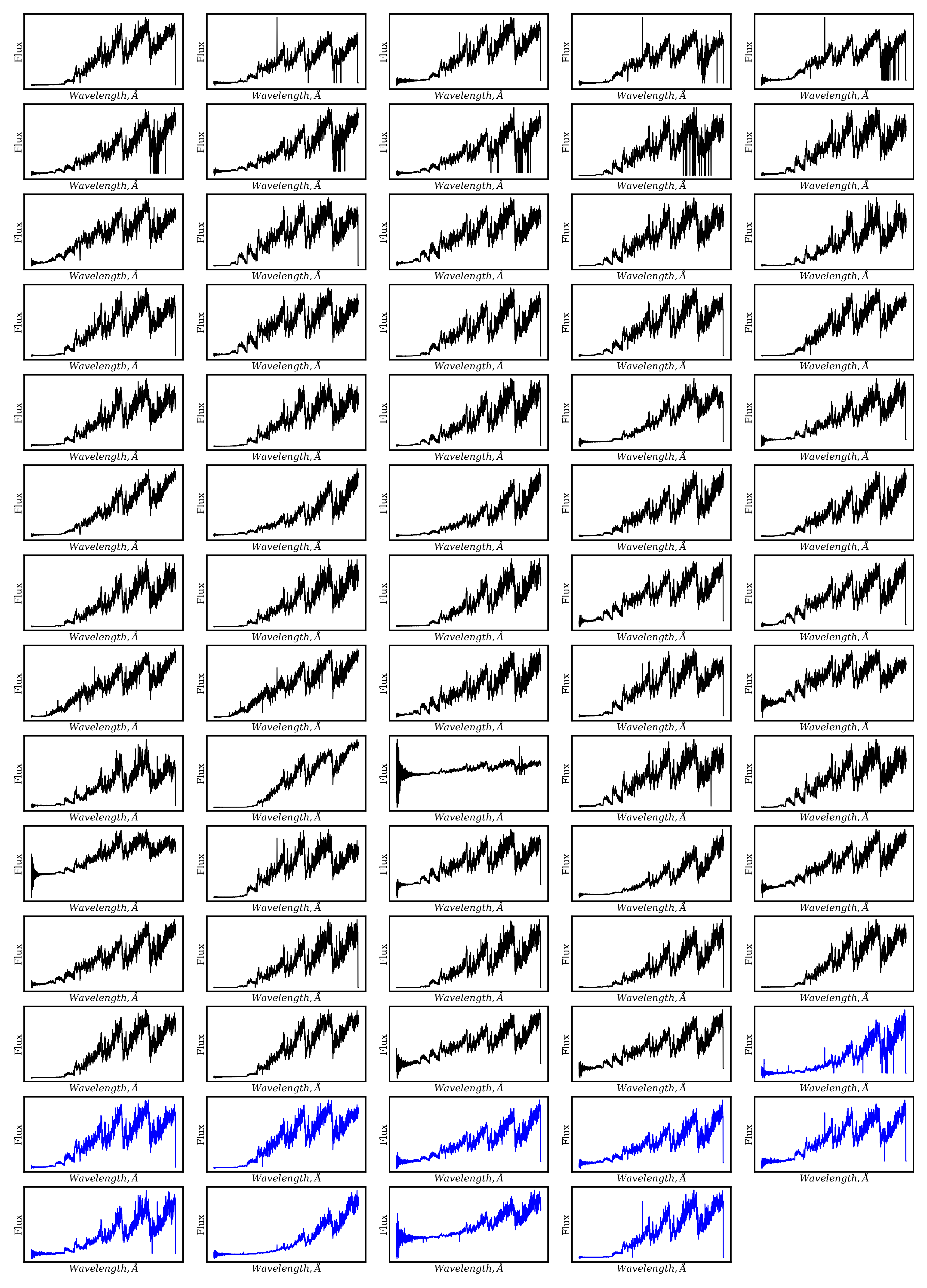}
	\caption{All 69 spectra for the 41 classical C-type stars identified in MaStar. Spectra used from the `goodspec' catalogue are plotted in black and `badspec' are in blue.}
    \label{fig:C-star_spectra-all}
\end{figure*}

\begin{figure*}
	\includegraphics[width=0.91\textwidth]{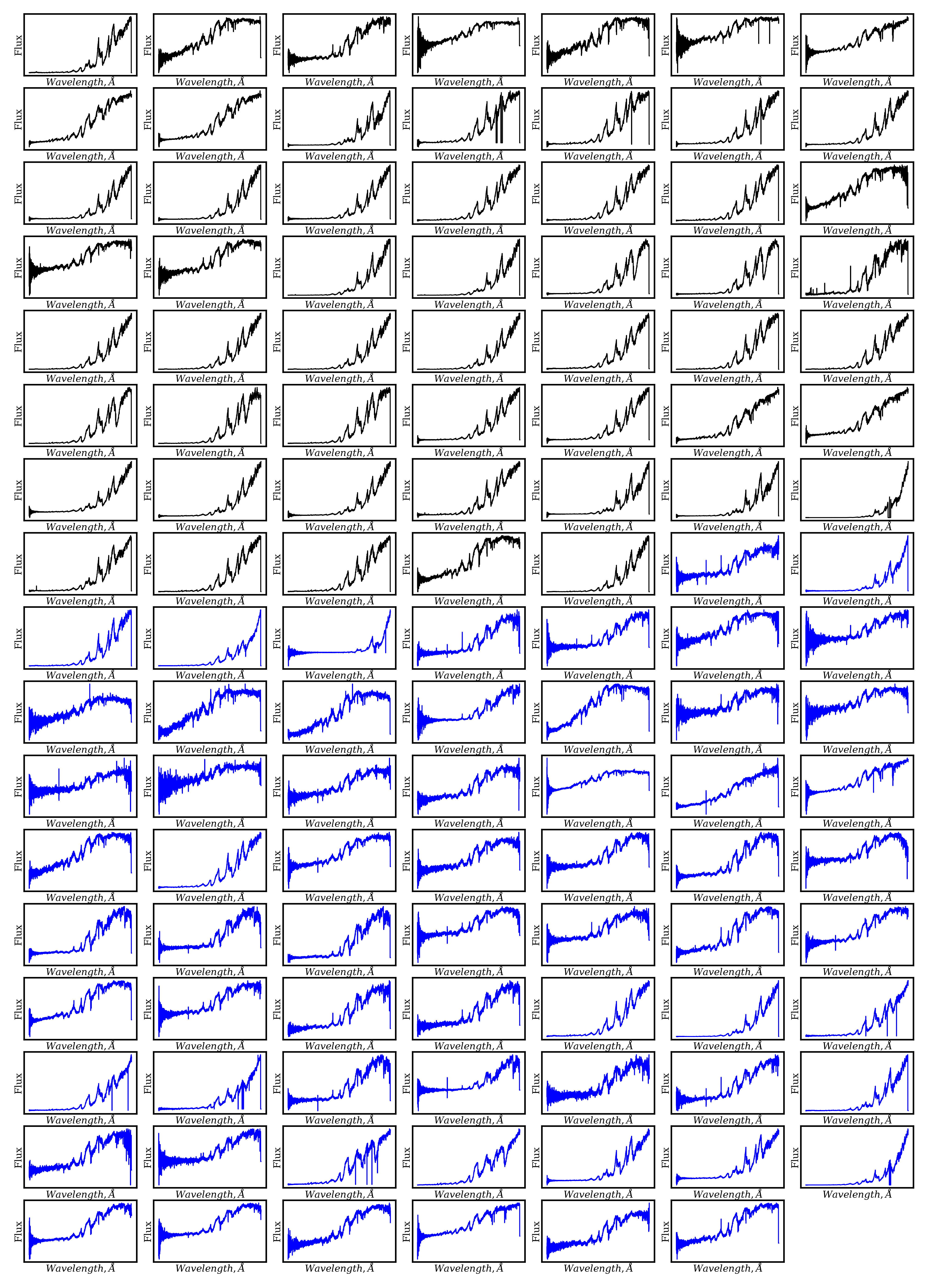}
	\caption{All 118 spectra for the 87 O-rich type stars identified in MaStar. Spectra used from the `goodspec' catalogue are plotted in black and `badspec' are in blue.}
    \label{fig:O-star_spectra-all}
\end{figure*}

\section{Parameters}
\onecolumn
\begin{landscape}
\begin{longtable}{|ccccccccccccccccc|}
\caption{Table showing various data for a sample of C- and O-rich type spectra. The order matches the spectra in Figures \ref{fig:C-star_spectra-all} and \ref{fig:O-star_spectra-all}, reading from left to right and top to bottom. The full table can be downloaded at \url{http://www.icg.port.ac.uk/mastar/}. Included is the MaNGA ID that represents each star in MaStar (1); MaStar plate number that holds fiber bundles (2); MaStar integral field unit (IFU) number that represents the fiber bundles that collected the spectrum (3); MaStar modified Julian date (MJD) number that represents the date of observation (4); median signal-to-noise ratio per pixel (5); sky position in degrees (6-7); colours used in our analysis (8-10); colours in the Johnson-Cousins system (11-13); whether the spectrum is from the badspec catalogue, where '1' is true and '0' is false (14); the spectral type (15); the bin number according to our spectral fitting using \citep{lancon_and_mouhcine_2002} and \citet{gonneau_etal_2017}, respectively (16-17). A value of -99 is used when no data is available.}
\label{table:params-all}
        \\\hline
        MaNGA ID & PLATE & IFU & MJD & S2N & RA & DEC & g - i & r - i & g - r & B - V & V - R & V - I & Badspec & Type & LM02 & XSL17 \\ 
        (1) & (2) & (3) & (4) & (5) & (6) & (7) & (8) & (9) & (10) & (11) & (12) & (13) & (14) & (15) & (16) & (17) \\ \hline
        55-1824778840680890240 & 10025 & 1901 & 58034 & 243.1 & 297.75 & 19.79 & 5 & 1.18 & 3.83 & -99 & -99 & -99 & 0 & C & 5 & 3 \\ \hline
        55-1824778840680890240 & 10025 & 1901 & 58243 & 113.26 & 297.75 & 19.79 & 3.88 & 0.82 & 3.05 & 3.35 & 1.47 & 2.49 & 0 & C & 1 & 3 \\ \hline
        55-1824778840680890240 & 10025 & 1901 & 58297 & 181.3 & 297.75 & 19.79 & 4.43 & 1.03 & 3.4 & 3.86 & 1.63 & 2.84 & 0 & C & 2 & 3 \\ \hline
        55-2191637904670592896 & 10092 & 1901 & 59034 & 165.4 & 321.05 & 59.46 & 3.94 & 0.85 & 3.09 & 3.57 & 1.41 & 2.45 & 0 & C & 1 & 3 \\ \hline
        55-2191637904670592896 & 10092 & 1901 & 59039 & 139.31 & 321.05 & 59.46 & 3.86 & 0.8 & 3.06 & 3.48 & 1.39 & 2.4 & 0 & C & 4 & 3 \\ \hline
        55-2179423086398277760 & 10092 & 3703 & 59034 & 137.31 & 324 & 59.14 & 3.44 & 0.97 & 2.47 & 2.66 & 1.4 & 2.57 & 0 & C & 3 & 3 \\ \hline
        55-2179423086398277760 & 10092 & 3703 & 59039 & 144.56 & 324 & 59.14 & 3.44 & 0.98 & 2.46 & 2.69 & 1.39 & 2.58 & 0 & C & 1 & 3 \\ \hline
        55-2179423086398277760 & 10092 & 3703 & 59067 & 136.78 & 324 & 59.14 & 3.38 & 0.94 & 2.44 & 2.76 & 1.35 & 2.51 & 0 & C & 1 & 3 \\ \hline
        55-1974906501932293632 & 10102 & 1902 & 58084 & 233.58 & 329.77 & 45.65 & 3.55 & 0.98 & 2.56 & 2.97 & 1.4 & 2.57 & 0 & C & 2 & 3 \\ \hline
        55-1974906501932293632 & 10102 & 1902 & 58087 & 149.15 & 329.77 & 45.65 & 3.4 & 0.93 & 2.47 & 2.92 & 1.35 & 2.46 & 0 & C & 2 & 3 \\ \hline
        60-2005409948097585920 & 11607 & 6104 & 58467 & 77.97 & 333.12 & 54.21 & 3.36 & 0.91 & 2.45 & 2.66 & 1.33 & 2.45 & 0 & C & 4 & 3 \\ \hline
        60-150975389014887424 & 12147 & 3701 & 58803 & 460.58 & 67.08 & 25.53 & 2.97 & 0.82 & 2.15 & 2.49 & 1.25 & 2.26 & 0 & C & 2 & 2 \\ \hline
        60-146753848479844992 & 12149 & 3703 & 58887 & 197.28 & 69.23 & 23.62 & 2.92 & 0.81 & 2.11 & 2.4 & 1.23 & 2.24 & 0 & C & 2 & 3 \\ \hline
        60-3411054853162179456 & 12152 & 3703 & 58853 & 304.21 & 70.98 & 20.15 & 3.35 & 0.98 & 2.37 & 2.71 & 1.37 & 2.53 & 0 & C & 2 & 3 \\ \hline
        60-3405514654587957760 & 12156 & 1902 & 58887 & 205.25 & 74.72 & 16.91 & 4.27 & 1.23 & 3.03 & -99 & -99 & -99 & 0 & C & 2 & 2 \\ \hline
        
        7-16889654 & 9295 & 6101 & 57941 & 48.71 & 283.6 & 16.76 & 4.83 & 2.67 & 2.16 & 1.36 & 1.92 & 4.71 & 0 & O & 7 & -99 \\ \hline
        3-58055914 & 10365 & 6101 & 58149 & 28.62 & 71.07 & 5.49 & 3.28 & 1.24 & 2.05 & 2.32 & 1.18 & 2.66 & 0 & O & 2 & -99 \\ \hline
        3-151441600 & 10388 & 1902 & 58236 & 6.33 & 122.85 & 33.94 & 4.37 & 2.22 & 2.15 & 2.61 & 1.49 & 3.94 & 0 & O & 3 & -99 \\ \hline
        3-160961632 & 10580 & 6101 & 58179 & 28.92 & 229.81 & 85.18 & 3.45 & 1.21 & 2.24 & 3.03 & 1.22 & 2.68 & 0 & O & 2 & -99 \\ \hline
        3-141269885 & 10637 & 3701 & 58179 & 15.22 & 157.29 & 43.46 & 3.28 & 1.25 & 2.03 & 2.55 & 1.12 & 2.67 & 0 & O & 2 & -99 \\ \hline
        3-111785612 & 11128 & 1902 & 58621 & 13.73 & 175.53 & 14.29 & 3.84 & 1.42 & 2.41 & -99 & 1.14 & 2.82 & 0 & O & 3 & -99 \\ \hline
        7-14732579 & 11256 & 3704 & 58567 & 58.65 & 260.44 & -25.29 & 3.85 & 1.63 & 2.22 & 2.47 & 1.42 & 3.23 & 0 & O & 3 & -99 \\ \hline
        60-4074329259731904512 & 11447 & 3703 & 58620 & 71.9 & 286.31 & -24.29 & 4.04 & 1.88 & 2.16 & 2.26 & 1.48 & 3.55 & 0 & O & 3 & -99 \\ \hline
        60-4074329259731904512 & 11447 & 3703 & 58622 & 70.97 & 286.31 & -24.29 & 4.05 & 1.84 & 2.21 & 2.31 & 1.48 & 3.52 & 0 & O & 3 & -99 \\ \hline
        60-6758882388961609344 & 11449 & 9101 & 58659 & 42.04 & 287.95 & -30.38 & 6.28 & 3.06 & 3.22 & 2.57 & 2.53 & 5.66 & 0 & O & 9 & -99 \\ \hline
        60-6759846450194240640 & 11450 & 9101 & 58605 & 93.68 & 289.48 & -27.97 & 4.69 & 2.46 & 2.24 & 1.75 & 1.78 & 4.36 & 0 & O & 4 & -99 \\ \hline
        60-6759846450194240640 & 11450 & 9101 & 58608 & 131.71 & 289.48 & -27.97 & 4.72 & 2.44 & 2.28 & 1.77 & 1.8 & 4.36 & 0 & O & 6 & -99 \\ \hline
        60-6759594009198666752 & 11450 & 9102 & 58605 & 126.07 & 290.42 & -28.36 & 4.41 & 2.32 & 2.09 & 1.75 & 1.63 & 4.12 & 0 & O & 6 & -99 \\ \hline
        
\end{longtable}
\end{landscape}